\documentclass[12pt]{article}
\usepackage{graphicx}
\usepackage{amsmath}
\usepackage{amssymb}
\usepackage{setspace}
\usepackage{dsfont}

\usepackage[round,comma]{natbib}

\graphicspath{{Plots/}}
\newcommand{\disc}{\boldsymbol\eta}

\newcommand{\x}{\textbf{x}}

\newcommand{\err}{\textbf{e}}

\newcommand{\obs}{\textbf{z}}

\newcommand{\sys}{\textbf{y}}

\newcommand{\best}{\textbf{x}^{*}}
\newcommand{\ens}{\textbf{F}}
\newcommand{\ensc}{\textbf{F}_{\boldsymbol{\mu}}}
\newcommand{\bas}{\boldsymbol\Gamma}
\newcommand{\bass}{\boldsymbol\gamma}

\newcommand{\rot}{\boldsymbol\Lambda}

\newcommand{\var}{\boldsymbol\Sigma}
\newcommand{\weight}{\textbf{W}}
\newcommand{\B}{\textbf{B}}

\newcommand{\cc}{\textbf{c}}

\newcommand{\R}{\mathcal{R}}

\newcommand{\fimpl}{\mathcal{I}}

\newcommand{\eps}{\boldsymbol{\epsilon}}

\newcommand{\Tau}{\boldsymbol{\mathcal{T}}}
\newcommand{\T}{\textbf{T}}
\newcommand{\s}{\textbf{s}}
\renewcommand{\t}{\textbf{t}}
\renewcommand{\c}{\textbf{c}}
\newcommand{\bmu}{\boldsymbol{\mu}}
\newcommand{\tspace}{\mathcal{X}}
\newcommand{\C}{\mathcal{C}}
\newcommand{\1}{\boldsymbol{1}}

\begin{document}

\title{Quantifying spatio-temporal boundary condition uncertainty for the North American deglaciation}

\author{James M. Salter$^1$\thanks{The authors gratefully acknowledge support from the EPSRC-funded Past Earth Network (Grant number EP/M008363/1). The authors would also like to thank the Isaac Newton Institute for Mathematical Sciences, Cambridge, for support and hospitality during the Uncertainty Quantification programme where work on this paper was undertaken (EPSRC grant no EP/K032208/1).}, Daniel B. Williamson$^{1,2}$, \\ Lauren J. Gregoire$^3$, Tamsin L. Edwards$^4$ \\ $^1$Department of Mathematics, University of Exeter, Exeter, UK. \\$^2$Alan Turing Institute, London, UK. \\ $^3$School of Earth and Environment, University of Leeds, Leeds, UK. \\ $^4$Department of Geography, King's College London, London, UK.}

\maketitle

\begin{abstract}
Ice sheet models are used to study the deglaciation of North America at the end of the last ice age (past 21,000 years), so that we might understand whether and how existing ice sheets may reduce or disappear under climate change. Though ice sheet models have a few parameters controlling physical behaviour of the ice mass, they also require boundary conditions for climate (spatio-temporal fields of temperature and precipitation, typically on regular grids and at monthly intervals). The behaviour of the ice sheet is highly sensitive to these fields, and there is relatively little data from geological records to constrain them as the land was covered with ice. We develop a methodology for generating a range of plausible boundary conditions, using a low-dimensional basis representation of the spatio-temporal input. We derive this basis by combining key patterns, extracted from a small ensemble of climate model simulations of the deglaciation, with sparse spatio-temporal observations. By jointly varying the ice sheet parameters and basis vector coefficients, we run ensembles of the Glimmer ice sheet model that simultaneously explore both climate and ice sheet model uncertainties. We use these to calibrate the ice sheet physics and boundary conditions for Glimmer, by ruling out regions of the joint coefficient and parameter space via history matching. We use binary ice/no ice observations from reconstructions of past ice sheet margin position to constrain this space by introducing a novel metric for history matching to binary data. 
\end{abstract}

\section{Introduction}

The last deglaciation, involving the melting of the North American ice sheet, occurred from the last glacial maximum around 21 thousand years ago (ka) onwards \citep{carlson2012ice}. By 6 ka, the ice sheet had almost disappeared from North America. The feedbacks between past climate and ice sheet melt are poorly understood \citep{ivanovic2016transient,ivanovic2018climatic}, with uncertainty in how much sea level rise can be attributed to ice melt caused by rapid warming events \citep{carlson2012ice}, so ice sheet models are used to study the deglaciation (e.g. \citet{gregoire2012deglacial,gregoire2016abrupt, patton2017deglaciation}). If the mechanisms that led to rapid warming and ice sheet melt in the past could be better understood, then these might be used to constrain predictions of future climate and ice sheet changes, improving their accuracy and reducing their uncertainty. However, climate is the largest source of uncertainty in modelling past ice sheet evolution \citep{seguinot2014effect,charbit2007numerical} and this source of uncertainty is challenging to characterise. 
\par
Ice sheet models are a set of partial differential equations (PDEs), containing parametrisations of physical processes that control the evolution of ice sheets. Thus ice sheet models include a number of parameters that control the flow and melt of ice sheets, but are only loosely constrained by observations. These parameters can be varied, with the output of the model being the evolution of the ice sheet extent, thickness and flow over time. As well as these parameters that control the ice sheet behaviour, simulating ice sheet evolution requires information about the climate, which controls the surface mass balance of an ice sheet (the balance between accumulation of snow, and melting of snow and ice, at the surface of the ice sheet). Although surface mass balance depends on processes that occur at small spatial and temporal scales (hourly, 1km or less) \citep{noel2016daily}, ice sheet models typically parameterise it as a function of monthly mean temperature and precipitation (e.g. the positive degree day method) \citep{reeh1989parameterization}. Such parametrisations are useful for simulating the evolution of large continental scale ice sheets over long time scales. Yet, the deglaciation occurs over millennia, so the monthly mean temperature and precipitation fields (the `boundary condition' to the ice sheet model) have an extremely high dimension: for a $48\times37$ spatial field (resolution of FAMOUS \citep{smith2008description}) varying monthly for 15,000 years (we focus here on 21 ka to 6 ka), this requires around 320 million values, for both temperature and precipitation separately.
\par
In order to have confidence in the model, the inputs must be tuned (`calibrated') so that the output matches historical observations (as Palaeo quantities cannot be directly measured, we use the term `observations' to refer to reconstructions obtained from proxies) of the ice sheet \citep{hourdinetal16}. To successfully calibrate such a computer model, it is important to vary both the input parameters and boundary conditions. Due to the dimension of the boundary conditions needed for ice sheet models, it is attractive to use the output of climate models \citep{gregoire2016abrupt}.
However, to run even a low-resolution climate model for the entire deglaciation is very expensive, requiring supercomputer time. Even when such runs are available, the output often does not match temperature records. Accurate past climate boundary conditions are needed to achieve a realistic deglaciation, and biases in the global climate model output may result in the ice sheet model being unable to reproduce historical observations of the ice sheet \citep{charbit2007numerical,seguinot2014effect}. 
The space of potential boundary conditions has an extremely high dimension, with the limited number of climate model runs forming only a small subset of this space, not necessarily containing the part of space that is consistent with temperature records. Therefore, an alternative method is required to properly evaluate the effect of uncertain boundary conditions on the model output.
\par
To tune computationally expensive models, the uncertainty quantification (UQ) field uses probabilistic calibration \citep{kennedy2001bayesian, higdon2008computer} and history matching \citep{craig1996bayes, craig2001bayesian}, using statistical models (`emulators') that can be evaluated quickly in place of the expensive computer model. Given the output of the computer model at a small number of settings of the inputs, an emulator is used to predict the output at unseen parameter settings, with an uncertainty on the prediction. Emulating and then calibrating computer models has been performed extensively \citep{williamson2013history, chang2014probabilistic,holden2015emulation,salter2018uncertainty,edwards2018}, but the uncertainty due to the boundary conditions is not always considered. \citet{pollard2016} and \citet{chang2016calibrating} apply emulation and calibration to an ice sheet model, varying input parameters relating to the ice sheet physics, but fixing the boundary condition using the output of another climate model.
\par
Methods for reducing high-dimensional input spaces have been developed, e.g. \citet{liu2017dimension} model the spatial input (bathymetry) of a tsunami model via a stochastic partial differential equation model, before using gradient-based kernel dimension reduction prior to building emulators. The bathymetry has 3200 dimensions, orders of magnitude smaller than the boundary conditions required for the deglaciation, with the bathymetry observed at locations across the whole spatial domain. In our application, the geological temperature observations are sparse, both spatially and temporally, with no observations over North America, the region for which we are studying the evolution of the ice sheet. The majority of observations are hundreds of kilometres apart, and ocean-based, so that an attempt to use a purely stochastic process-based model for the boundary conditions may have problems setting appropriate correlations between the sparse observations, and overcoming the biases caused by this inhomogeneous spatial distribution, with some regions having few records (e.g. interior of continents).
\par
In this paper, we develop a novel method that enables us to define more plausible boundary conditions, using a low-dimensional representation of the full boundary condition input. This method exploits physical spatio-temporal structure in existing low-resolution climate model ensembles, whilst retaining enough flexibility to overcome the biases in these models. By varying a small number of coefficients that control the basis representation, we efficiently generate past climates that are consistent with observations. We can then better explore the uncertainty in the ice sheet model output, by jointly varying the ice sheet parameters and the boundary conditions, then searching for combinations that lead to output consistent with ice sheet observations using emulation and calibration (history matching).
\par
Section \ref{icesheetmodel} provides an overview of the Glimmer ice sheet model we use to simulate the North American ice sheet, and of the types of palaeo-data that are available. Section \ref{emulationhm} outlines the statistical methods from the UQ literature we use to analyse the output of Glimmer. Section \ref{modelsection} gives our framework for modelling and calibrating boundary conditions, with the boundary condition model fitted in Section \ref{applicationsection}. Section \ref{definec} defines a bounded space for the boundary conditions, and Section \ref{binarysection} provides additional history matching methodology required to compare ice sheet thickness with binary observations. Section \ref{hmresults} shows how boundary condition uncertainty for Glimmer is reduced by history matching, and Section \ref{discussion} provides discussion.

\section{The Glimmer ice sheet model and Palaeo observations} \label{icesheetmodel}

Glimmer is a fast three-dimensional ice sheet model \citep{rutt2009glimmer} which has been used to simulate the past evolution of the North American ice sheet \citep{gregoire2012deglacial} (\citet{mcneall2013potential} use Glimmer to give output for the Greenland ice sheet instead). The model simulations are set up as in \citet{gregoire2016abrupt}. The output resolution can be adjusted; here, we have a $194 \times 150$ grid (40 km resolution) covering the North American ice sheet.  There are numerous uncertain parameters in Glimmer that control the flow and melt of the ice sheet. Previous work has identified 7 parameters, here referred to as $\x$, that have the strongest effect on the output of the model, controlling aspects of the ice sheet such as basal sliding and lapse rate \citep{hebeler2008impact,gregoire2010modelling,gregoire2016abrupt}. The model uses a positive degree day (PDD) scheme to simulate ice sheet surface mass balance, driven by monthly mean temperature and precipitation fields input every month between 21 ka and 6 ka, covering the retreat of the North American ice sheet.
\par
\begin{figure}[t]
\centering
\includegraphics[width = 0.55\linewidth]{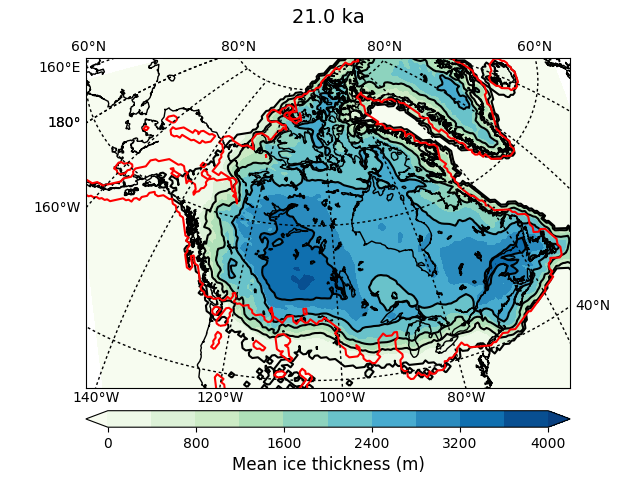}
\caption{An example of the ice thickness output given by Glimmer at the start of the deglaciation (21 ka). The red line shows the observed extent of the ice sheet.}
\label{glimmerexample}
\end{figure}

Figure \ref{glimmerexample} gives an example of the spatial output of Glimmer, at 21 ka, with darker blue representing a thicker ice sheet, and the observed extent of the ice sheet (the ice margin) given by the red line (described in Section \ref{iceobssection}).
\par
For the boundary condition input of Glimmer, due to its extremely high dimension, the output of a transient run of a low-resolution global climate model (GCM), for example FAMOUS \citep{smith2008description}, is typically used to provide a physically-coherent boundary condition.
\par
There are several problems with using GCM output as the Glimmer boundary condition. There are only a limited number of modelling groups who have been able to simulate the last deglaciation with climate models. Two groups have used intermediate complexity climate models of fairly high resolution \citep{menviel2011deconstructing, roche2011deciphering}, whilst two other groups have used GCMs, with simulations taking from 4 months to 2 years to complete \citep{liu2009transient,gregoire2012deglacial}. Several more groups around the world have now coordinated to run a larger multi-model ensemble of the deglaciation within the Palaeoclimate Modelling Intercomparison Project phase 4 \citep{ivanovic2016transient} (data not available at time of writing).
\par
The GCMs have their own input parameters, which have not been tuned with temperatures through the deglaciation. For example, the simulations carried out with the CCSM3 model \citep{liu2009transient} are tuned to modern-day climate, while simulations with the FAMOUS GCM were tuned to simulate both modern and glacial climate, but not the period in-between \citep{gregoire2011optimal}. Therefore, the boundary conditions currently used for Glimmer do not adequately match geological observations, nor does their spread capture uncertainty in those boundary conditions.
\par
\begin{figure}[t]
\centering
\includegraphics[width = 0.75\linewidth]{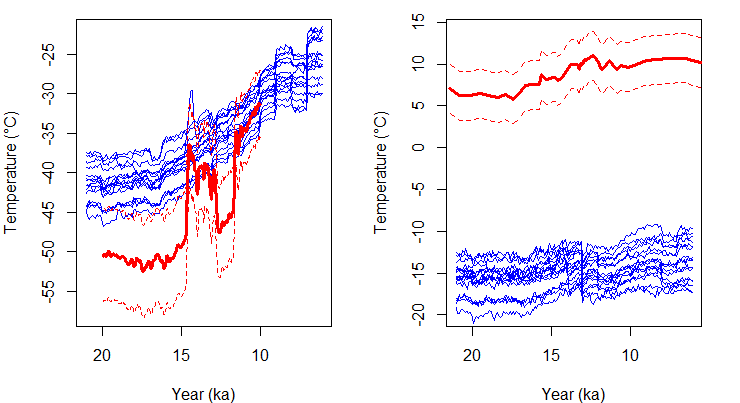}
\caption{The time series of temperature for each of the 16 ensemble members (blue) and the observed temperatures (red), for Greenland (left) and Alaska (right), with 95\% observation error uncertainty given by the dotted lines. Some of the abrupt changes in temperature seen for every ensemble member are due to the climate model's ice sheet topography being updated every 1,000 years.}
\label{GreenAlaskaObs}
\end{figure}

To illustrate this difference between the model and geological observations, we consider an ensemble of 16 GCM runs that have been used as boundary conditions for Glimmer: 15 FAMOUS simulations \citep{gregoire2011optimal}, and 1 TRACE simulation \citep{liu2009transient}. Each ensemble member consists of monthly temperature and precipitation fields from 21 ka onwards, with FAMOUS on a $48\times37$ spatial grid ($7.5^{\circ}\times3.5^{\circ}$), and TRACE a $96\times48$ grid (3.75$^{\circ}\times3.75^{\circ}$). Figure \ref{GreenAlaskaObs} compares the ensemble temperatures to observations in Greenland and Alaska (two of the closest observed spatial locations to North America).
This plot shows that the GCM output is biased away from the observations in important locations. In Greenland, the ensemble is too warm at the start of the deglaciation, with the majority of runs not falling within the 95\% error bounds on the observations. The runs generally fail to replicate the rapid warming around 14.7 ka, and none capture the rapid cooling around 13 ka. In Alaska, the temporal pattern is correct, but the ensemble is 15$^\circ$C cooler than the observations, a difference of 10 standard deviations from the observations.
\par
Running Glimmer with boundary conditions that exhibit biases such as those in Figure \ref{GreenAlaskaObs}, with important rapid warming and cooling periods not captured, and the temperature several degrees wrong in key locations, will lead to inaccurate simulations of the deglaciation, and difficulties in studying the effect that rapid warming and cooling events have on the ice sheet.

\subsection{Temperature observations} \label{tempobs}

In addition to the temperature observations for Greenland and Alaska, we have observations at other spatio-temporal locations, but none over the North American ice sheet \citep{shakun2012global,buizert2014greenland}.
\par
\begin{figure}[t]
\centering
\includegraphics[width = 0.65\linewidth]{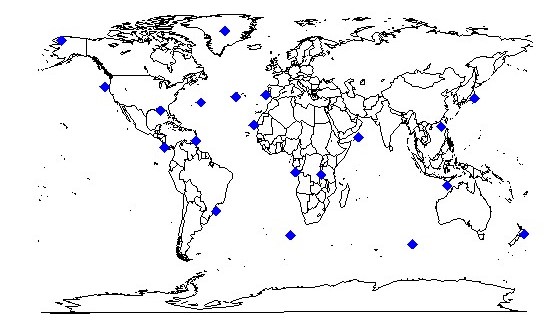}
\caption{Map showing the locations of observed temperatures.}
\label{obslocations}
\end{figure}

Figure \ref{obslocations} shows 20 spatial locations where temperature observations are available. These observations are reconstructed from various different sources, for example ice cores (Greenland), Mg/Ca, $U_{37}^{k'}$, and microfossils, and have varying degrees of uncertainty (see \citet{shakun2012global}): e.g., for microfossils, 1 standard deviation is given as 1.5$^{\circ}$C, as in Alaska, whilst 1 standard deviation for the Greenland ice core varies between 2.1 and 3.2 through time. \citet{shakun2012global} give additional spatial locations, but we use this subset, ignoring sources considered less reliable and those in close proximity to chosen observations. Our subset contains those observations closest to North America, whilst also having well-spaced points around the world.
\par
The geological records are also irregular in time: in the 20 locations in Figure \ref{obslocations}, the number of time points with an observation varies from only 26 to over 100 (e.g. the Alaska observations plotted in Figure \ref{GreenAlaskaObs} are based on 43 observed time points). These are infrequent compared to the number of time points in the boundary condition, $15,000\times12 = 180,000$. Furthermore, the observations have a temporal uncertainty, with this generally increasing as we move further back in time, where 1 standard deviation can be 400 years or more.

\subsection{Ice sheet observations} \label{iceobssection}

\begin{figure}[t]
\centering
\includegraphics[width = 0.45\linewidth]{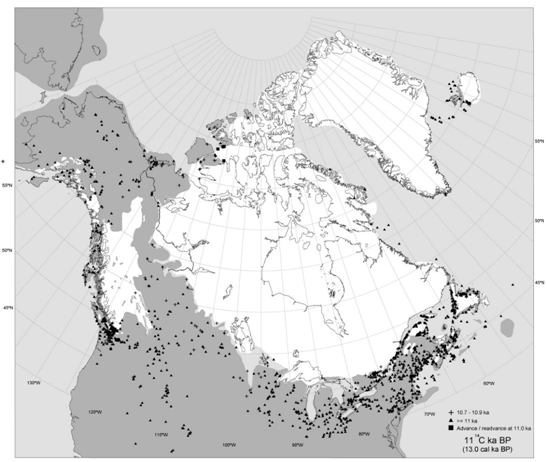}
\caption{The ice extent observation at 13 ka.}
\label{obsextent}
\end{figure}

Glimmer gives ice thickness (in metres) as an output, but there are no past observations of this. The most robust constraints on past ice sheet evolution are reconstructions of the extent through time. For the North American ice sheet, the latest reconstruction is from \citet{dyke2004outline}, an example of which is shown in Figure \ref{obsextent}. This dataset provides estimates of ice margin position at quasi-regular 500-1000 year intervals through the deglaciation, based on expert interpretation of compilations of geological data that date the presence or absence of ice. We therefore have a set of maps showing presence or absence of ice in each gridbox at given times throughout the deglaciation. 
\par
We also have estimates of the volume of the North American ice sheet through the deglaciation, inferred from ensembles of ice sheet models constrained with data on volume and extent \citep{tarasov2012data}. We use the spatio-temporal patterns of ice extent and the time series of ice volume to calibrate Glimmer within a history matching framework.

\section{Emulation and history matching} \label{emulationhm}

As well as having a high-dimensional input, Glimmer's output, $f$, is spatio-temporal, so we may require multivariate emulation methods to model $f(\x)$. We vectorise the output so that $f(\cdot)$ is a vector of length $\ell$, the number of outputs, and define an ensemble as $\ens = (f(\x_1), \ldots, f(\x_n))$, where design $\textbf{X} = (\x_1, \ldots, \x_n)$ consists of input parameter settings from space $\tspace$. Here, we treat any boundary condition parameters as part of $\x$.
\par
The most common method for emulating spatial output is via a low-dimensional basis representation, often the basis given by the singular value decomposition (SVD) of the (centred) model output \citep{higdon2008computer, wilkinson2010bayesian, sexton2011multivariate, chang2014probabilistic,salter2018uncertainty}:
\begin{equation} \label{svd}
\ensc^T = \textbf{U} \boldsymbol{\Sigma} \bas^T,
\end{equation} 
where the $i^{th}$ column of $\ensc$ is given by $f(\x_i) - \bmu$, for ensemble mean $\bmu$. The columns of $\bas$ form a basis for $\ensc$, and we project output onto this basis via
\begin{equation} \label{projecteqn}
\textbf{c}(\x) = (\bas^{T}\weight^{-1}\bas)^{-1} \bas^{T} \weight^{-1} (f(\x)  - \boldsymbol{\mu}),
\end{equation}
for positive definite variance matrix $\weight$. The original field is reconstructed as
\begin{equation} \label{reconeqn}
f(\x) = \bas \textbf{c}(\x) + \boldsymbol{\mu} + \boldsymbol{\epsilon},
\end{equation}
with error $\boldsymbol{\epsilon} = \textbf{0}$ for $\x \in \textbf{X}$, and $\cc(\x) = (c_1(\x), \ldots, c_n(\x))^T$.
\par
The first $q << n$ basis vectors explain the majority of the variability in $\ensc$, hence $\bas$ is truncated to give basis $\bas_q = (\bass_1, \ldots, \bass_q)$, where $\bass_i$ is the $i^{th}$ column of $\bas$. Univariate emulators (commonly Gaussian process-based) for the $q$ coefficients given by projection onto $\bas_q$ are constructed:
\begin{equation} \label{projectemulate}
c_i(\x) \sim \text{GP}(m_i(\x), K_i(\x, \x)), \quad i = 1, \ldots, q,
\end{equation}
for mean function $m_i(\cdot)$ and covariance function $K_i(\cdot, \cdot)$. 
\par
When emulating and calibrating a spatial field, we select an optimal basis via a rotation of the SVD (or weighted SVD) basis \citep{salter2018uncertainty}. This allows patterns from the observations, which may not be present in the truncated basis $\bas_q$, yet may appear as a linear combination of several low-eigenvalue vectors of $\bas$, to be incorporated into the calibration basis, ensuring that the correct directions of output space can be searched (i.e. so that the basis choice doesn't guarantee that we conclude that the model cannot represent the observations). The optimal rotation algorithm of \citet{salter2018uncertainty} finds a rotated basis, $\bas^* = \bas \rot$, by optimising
\begin{equation} \label{rotationminimise}
\min_{\rot} \R_{\weight}(\bas^*_{q^*}, \obs),
\end{equation}
where
\begin{align*}
\begin{split}
\R_{\weight}(\B, \obs) &= \lVert \obs - \textbf{r}(\obs) \rVert_{\weight} = (\obs - \textbf{r}(\obs))^{T} \weight^{-1} (\obs - \textbf{r}(\obs)), \\
\textbf{r}(\obs) &= \B (\B^{T} \weight^{-1} \B)^{-1} \B^{T} \weight^{-1} \obs,
\end{split}
\end{align*}
for positive definite $\weight$, basis $\B$ and `reconstruction error' $\R_{\weight}(\B, \obs)$, representing the difference between the observations, $\obs$, and their reconstruction, $\textbf{r}(\obs)$, given by the subspace defined by $\B$. The optimisation in \eqref{rotationminimise} is performed subject to constraints that ensure there is enough ensemble signal on each basis vector, so that emulators can be built for the coefficients given by projection onto the first $q^*$ basis vectors. For further discussion and code, see \citet{salter2018uncertainty}.
\par
Due to the extremely high dimension of $\x$ in our application, it is not possible to automatically apply one of the above methods. We return to this problem in Section \ref{modelsection}.

\subsection{History matching} \label{sectionhm}

History matching is a method for calibrating the input parameters of a computer model, removing regions of parameter space that are unlikely to lead to output consistent with observations \citep{craig1996bayes, vernon2010galaxy}. Given observations, $\obs$, of a physical system, $\sys$, that is represented by computer model $f(\cdot)$, we assume that
\begin{equation} \label{hmequation}
\sys = f(\best) + \disc, \quad  \obs = \sys + \err,
\end{equation}
where $\err \sim \mathrm{N}(\textbf{0}, \var_{\err})$ is the observation error, $\disc \sim \mathrm{N}(\textbf{0}, \var_{\disc})$ is the discrepancy between the `best' input of the computer model, $\best$, and the true system, and the terms in \eqref{hmequation} are independent  \citep{kennedy2001bayesian}. Given $\var_{\err}$, $\var_{\disc}$, and an emulator for $f(\x)$, the implausibility of $\x$ is
\begin{equation} \label{impl}
\mathcal{I}(\x) = (\obs - \text{E}[f(\x)])^{T}(\text{Var}(\obs - \text{E}[f(\x)]))^{-1} (\obs - \text{E}[f(\x)]).
\end{equation}
The space of not implausible runs (`not ruled out yet' (NROY) space) is
\begin{equation} \label{nroyequation}
\mathcal{X}_{NROY} = \{ \x \in \mathcal{X} | \mathcal{I}(\x) < b \},
\end{equation}
for a bound, $b$, used to rule out implausible settings of $\x$. For univariate $f(\cdot)$, Pukelsheim's 3-sigma rule \citep{pukelsheim1994three} is used; for multivariate $f(\cdot)$, $b$ is often set as the 99.5\% value of the chi-squared distribution with $\ell$ degrees of freedom \citep{vernon2010galaxy}. By designing new ensembles within NROY space, history matching can be performed iteratively over multiple waves (`refocussing') \citep{vernon2010galaxy, salter2016comparison, williamson2017tuning}. 
\par
The output of Glimmer is ice thickness, which is not directly comparable to the binary observations of ice extent, hence we require an extension to the standard history matching methodology, addressed in Section \ref{binarysection}.

\section{A framework for calibrating boundary conditions} \label{modelsection}

We now introduce our novel framework for modelling and calibrating spatio-temporal boundary conditions with sparse observations, as in Paleao climate problems.
\par
Our approach consists of representing the high-dimensional boundary condition input of Glimmer using a low-dimensional basis representation, controlled by a small number of coefficients, the vectors of which we select sequentially - temporally then spatially. By varying these coefficients, which can be treated as additional input parameters in a traditional calibration exercise, we calibrate both the boundary condition and the parameters that control ice sheet physics.
\par
Our specific problem is characterised by a small computer model ensemble and limited observations. It is possible to fit a spatio-temporal Gaussian Markov random field to the observations using INLA \citep{liu2017dimension}, varying the parameters of this process to generate boundary conditions. However, such a process will revert towards its prior mean as the distance increases from an observed point, a problem here due to the extreme sparsity of the observations. The majority of observations are ocean-based, so that the Gaussian process model may also be biased over continents, hence defining a structured mean function is important. We use the GCM output to give our model physical structure, fitting it in such a way that can better capture the observations.

\subsection{Modelling the boundary condition}

Let $\T$, a vector with length $\ell_s\ell_t$, where $\ell_s$ and $\ell_t$ are the number of spatial and temporal dimensions respectively, represent a spatio-temporal boundary condition (e.g. for temperature). We have $n$ runs of the climate model $\tau(\cdot)$, which would normally be used to force Glimmer explicitly, giving ensemble $\Tau = (\tau(\x_1), \ldots, \tau(\x_n))$, the mean of which is $\bmu$. We model $\T$ as:
\begin{equation}\label{boundaryModel}
\T \mid \c \sim \mathrm{MVN}(h(\c), \var_s\otimes\var_t), \qquad h(\c) = \bmu + \sum_{j=1}^{n_t}c^t_j\t_{j} + \sum_{j=1}^{n_s}c^s_j\s_j,
\end{equation}
with $\c = (\c^t,\c^s)$ a vector of coefficients that controls the boundary condition, and $\t$ and $\s$ sets of $(\ell_s\ell_t)$-dimensional basis vectors that capture the principal temporal variability ($\t$) and spatial variability ($\s$) in $\Tau$. We describe selecting $\t$ and $\s$ in Section \ref{sectionbasisselection}. The coefficients, $\c \in \C$, are the set of calibration parameters for our boundary condition, $\T$, with the space $\C$ defined in Section \ref{definec}.  This parametrised model allows boundary condition uncertainty to be explored, by jointly varying $\cc$ and inputs $\x$, and running Glimmer at these choices. The basis vectors will be constructed so that $n_t + n_s$ is small, allowing the joint ice sheet parameter - coefficient input space to be efficiently explored, whilst having enough flexibility so that $\T$ is able to represent observations.
\par
We model the sparse observations of the boundary condition as:
\begin{equation} \label{obsModel}
\obs_T \sim \mathrm{MVN}(\T, \var_s\otimes\var_{t'}),
\end{equation}
with the observed entries of this vector denoted by $\obs_{T,obs}$. In \eqref{boundaryModel} and \eqref{obsModel}, we have assumed a Kronecker variance structure, with a common spatial variance matrix, for computational convenience (although $\var_t$ and $\var_{t'}$ may be different).

\subsection{Simulating boundary conditions} \label{completeprocess}

We generate $\T$ by choosing $\c$, and taking the expectation of $\pi(\T \mid \obs_{T, obs}, \c)$. This distribution is found via:
\begin{align} \label{BCupdate}
\begin{split}
\pi(\T \mid \obs_{T, obs}, \c) &= \int \pi(\T, \obs_{T, miss} \mid \obs_{T, obs}, \c) d \obs_{T, miss} \\
&= \int \pi(\T \mid \obs_{T, miss}, \obs_{T, obs}, \c) \pi(\obs_{T, miss} \mid \obs_{T, obs}, \c) d \obs_{T, miss}.
\end{split}
\end{align}
This is tractable using the Kronecker structure on the variance in equations \eqref{boundaryModel} and \eqref{obsModel}. Without the Kronecker assumption, conditioning the full spatio-temporal boundary condition on observations would involve the inversion of an $\ell_s\ell_t\times\ell_s\ell_t$ matrix (in our application, $\ell_s = 48\times37$, and $\ell_t = 15000\times12$), whereas matrices with a Kronecker structure can be efficiently inverted via $(\var_s \otimes  \var_t)^{-1} = \var_s^{-1} \otimes \var_t^{-1}$. We integrate over the missing entries, $\obs_{T,miss}$, of the sparse observation vector, because conditioning on only $\obs_{T,obs}$ would require the variance matrix to be expanded, losing the Kronecker structure that makes this calculation tractable. Additionally, without either a common spatial or temporal variance matrix across equations \eqref{boundaryModel} and \eqref{obsModel}, the Kronecker structure would be lost (as we need to sum the variance matrices prior to inversion, see Section \ref{smcalculation}).
\par
The final boundary condition at $\c$ is given by the expectation of \eqref{BCupdate} (derived in Section \ref{smcalculation}):
\begin{displaymath}
\text{E}[\T \mid \obs_T, \cc] = h(\cc) + vec(\var_t (\var_{t'} + \var_t)^{-1} (\obs_T - h(\cc))^T (\var_{s} \var_{s}^{-1})^T).
\end{displaymath} 
We could instead sample from \eqref{BCupdate}, but use the expectation here so that deterministic emulators can be built.

\subsection{Selecting basis vectors} \label{sectionbasisselection}

The previous section has given the general model structure. Now, we provide a method for finding the set of basis vectors, $(\t, \s)$. We use a sequential process, rather than aiming to directly select spatio-temporal basis vectors, or all of the vectors at once, to aid model flexibility and interpretability of the resulting basis.
\par
Given $\Tau$ and $\obs_{T,obs}$, we first select a small number, $n_t$, of linear combinations of the climate model output that explain the key temporal features of the boundary conditions, and then adjust for any spatial biases, in important regions for the deglaciation.  Calculating the SVD basis across the full spatio-temporal ensemble is expensive, and restricts the potential basis vectors we can find by forcing correlations that may not exist in reality, due to the limited number of degrees of freedom in $\Tau$. Our approach allows more variability to be captured by partitioning $\Tau$, whilst maintaining physical interpretability: it is much clearer what effect a spatial or temporal coefficient has on the resulting boundary condition, rather than one that controls a spatio-temporal field.
\par
We desire basis vectors that allow the observed temperatures to be matched more accurately (compared to $\Tau$). As the SVD basis itself may not directly provide vectors that allow the observations to be captured due to its variance-maximising property, we will consider all basis vectors in the subspace defined by the climate ensemble, and find those that represent observations as well as possible, using the basis rotation algorithm from \citet{salter2018uncertainty} to aid construction of the basis. By restricting ourselves to the space defined by the climate ensemble, we are able to preserve many of the physical spatio-temporal relationships given by the climate model. By not simply using the leading SVD basis vectors, we are able to reproduce a less homogeneous set of boundary conditions than the ensemble.

\subsubsection{Temporal basis vectors} \label{temporalvectors}

We first centre the ensemble by its mean, 
\begin{displaymath}
\Tau_{\bmu} = \Tau - \bmu = (\tau(\x_1) - \bmu, \ldots, \tau(\x_n) - \bmu) = (\tau_{\bmu}(\x_1), \ldots, \tau_{\bmu}(\x_n)),
\end{displaymath}
and extract the main temporal variability from $\Tau_{\bmu}$. From the set of observed spatial locations, $\mathcal{S}$, we select a subset $\mathcal{S}_t \subset \mathcal{S}$ of $n_{\mathcal{S}_t}$ locations, with which to fit the temporal basis vectors, $\t$. Let a subscript of $\mathcal{S}_t$ denote the selection of all entries corresponding to the spatial locations in $\mathcal{S}_t$, e.g. $\Tau_{\bmu, \mathcal{S}_t}$ is the centred ensemble restricted to $\mathcal{S}_t$, with dimension ($n_{\mathcal{S}_t} \times  \ell_t) \times n$.
\par
With this subset of the ensemble, we calculate the SVD basis $\bas$:
\begin{equation} \label{svdbasis}
\Tau_{\bmu, \mathcal{S}_t}^T = \textbf{U} \boldsymbol{\Sigma} \bas^T.
\end{equation}
In order to represent the observations as well as the ensemble allows for the locations in $\mathcal{S}_t$, we find a rotated basis $\bas^* = [\bas \rot]_{\cdot, 1:n_t}$ such that the difference between the observations and their basis reconstruction,
\begin{displaymath}
\R_{\weight}(\bas^*, \obs_{T, obs, \mathcal{S}_t} - \bmu_{\mathcal{S}_t}),
\end{displaymath}
is minimised. From \eqref{obsModel}, we have $\weight = [\var_s]_{\mathcal{S}_t} \otimes \var_{t'}$, the entries of the observation error matrix that correspond to $\mathcal{S}_t$. The rotated basis is truncated after the first $n_t$ basis vectors, where $n_t < n$ is enough to capture the majority (e.g. 95\%) of the variability in $\obs_{T, obs, \mathcal{S}_t}$.
\par
These basis vectors only relate to $\mathcal{S}_t$. To convert these into the full spatio-temporal vectors required by \eqref{boundaryModel}, we show (in Section \ref{smtemporal}) that taking a linear combination of these $n_t$ basis vectors, using $n_t$ coefficients, can be rewritten as a linear combination of the $n$ ensemble members, using the properties of the SVD basis. Setting:
\begin{displaymath}
\t_j = [\Tau_{\mu} (\textbf{U} \boldsymbol{\Sigma}^{-1} \rot)]_{\cdot j}, \quad j = 1, \ldots, n_t,
\end{displaymath}
gives the basis vectors in \eqref{boundaryModel}.
\par
By extrapolating to linear combinations of the spatio-temporal ensemble, we have retained some physicality and smoothness from the climate model output. The benefit of this method is that rather than specifying coefficients for the $n$ ensemble members, we only need $n_t$ (usually 2 or 3), whilst the coefficients remain interpretable, and the possible linear combinations are naturally restricted to the subspace of the ensemble that best matches observations for $\mathcal{S}_t$.

\subsubsection{Spatial basis vectors} \label{spatialvectors}

Matching to a small number of locations at the previous stage, and using these to infer full spatio-temporal vectors, may have induced biases in the output in important spatial locations. We also know that there were biases in the climate ensemble originally. The next step in our methodology is to add spatial basis vectors to correct any such spatial biases.
\par
We assume that we have captured as much of the temporal variability as possible, and remove the temporal aspect by averaging across time, only considering spatial patterns not accounted for in the ensemble mean or the temporal basis vectors. To do so, we centre the ensemble and observations by $\bmu$, and then remove their projection onto $(\t_1, \ldots, \t_{n_t})$. Let the coefficients given by projection of the observations, and of ensemble member $i$, be denoted by $\c_{\obs}$ and $\c_i$ respectively, then
\begin{align} \label{spatialresidual}
\begin{split}
\epsilon_{\Tau} &= (\epsilon_{\Tau,1}, \ldots, \epsilon_{\Tau,n}), \quad \epsilon_{\Tau,i} = \Tau_{\mu} - (\t_1 \ldots \t_{n_t}) \c_i, \\
\epsilon_{\obs} &= \obs_T - \bmu - (\t_1 \ldots \t_{n_t}) \c_{\obs}, \\
\end{split}
\end{align}
represent the spatial variability in the observations and climate ensemble that is yet to be explained, with the averages across time given by $\bar{\epsilon}_{\Tau}$, $\bar{\epsilon}_{\obs}$.
\par
As in Section \ref{temporalvectors}, we calculate the SVD basis of $\bar{\epsilon}_{\Tau}$. Denoting this basis $\bas_{\epsilon}$, we search for a rotation that best represents the remainder of the observations ($\bar{\epsilon}_{\obs}$), minimising:
\begin{displaymath}
\R_{\weight}((\bas_{\epsilon} \rot_{\epsilon})_{\mathcal{S}_s,1:n_s}, \bar{\epsilon}_{\obs, \mathcal{S}_s}),
\end{displaymath}
for rotation matrix $\rot_{\epsilon}$, $\weight = [\var_s]_{\mathcal{S}_s}$, $n_s$ the number of selected spatial basis vectors, and $\mathcal{S}_s$ a subset of the observed spatial locations. The rest of the locations are reserved for diagnostic checks, to ensure we are not overfitting to the observed locations, again aiming to maintain physicality from the climate model. To convert from spatial to spatio-temporal vectors, as required in the model \eqref{boundaryModel}, we then set:
\begin{displaymath}
\s_j = \1 \otimes (\bas_{\epsilon}\rot_{\epsilon})_{\cdot j}, \quad j = 1, \ldots, n_s,
\end{displaymath}
for $\1$ an $\ell_t$-vector of 1s, i.e. the chosen spatial vectors are repeated at every time point, giving the desired spatial corrections.

\subsection{Application to Glimmer} \label{applicationsection}

We fitted the boundary condition model to the geological temperature observations, to give a set of plausible past temperatures with which to run Glimmer. In this application, we do not explicitly fit a model for the precipitation boundary condition, instead using monthly mean precipitation from the standard FAMOUS simulation as in \citet{gregoire2012deglacial} for consistency with previous work.
\par
Based on discussions with experts, we fit separate models to give the flexibility required to capture the variability in the observations throughout the 15,000 years. The temporal domain is split into three (21 ka to 15 ka, 15 ka to 13 ka, and 13 ka to 6 ka), and a separate basis is fitted for each, with the short second interval containing the key rapid warming in Greenland. Glimmer is typically run by taking 100 year averages of GCM output, hence we average across 100 year periods instead of fitting the model using monthly GCM data, reducing the overall dimension of the boundary condition by a factor of 100, to around 3 million, and giving more consistency with the geological observations (typically on a scale of at least 100 years). We address smoothing between time periods and seasonality in Sections \ref{smoothing} and \ref{seasonality}.

\subsubsection{Fitting the temporal basis} \label{secfittingmodel}

The ensemble mean $\bmu$ gives an underlying spatio-temporal field. To construct $\t$, altering the temporal patterns in the model, we used Greenland as $\mathcal{S}_t$, as the observations in this location contain the rapid warming and cooling between 15 ka and 11 ka, not currently represented in the GCM-only boundary condition, that we believe it is important for our model to capture. Greenland is also one of the closest spatial locations to the North American ice sheet. 
\par
\begin{figure}[t]
\centering
\includegraphics[width=0.75\linewidth]{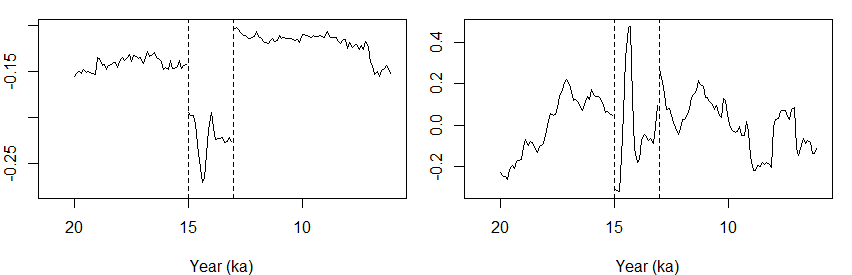}
\caption{The temporal basis vectors chosen using Greenland, from which we can calculate coefficients for the (centred) ensemble members. Left: first basis vector for each time period, right: second basis vector for each time period.}
\label{temporalbasis}
\end{figure}

For each of the time periods, a single basis vector was not sufficient to represent the observed time series in Greenland. However, adding a second allowed the main patterns to be captured, with little additional benefit found by adding a third, resulting in basis vectors $\t = (\t_{1,1}, \t_{2,1}, \t_{1,2}, \t_{2,2}, \t_{1,3}, \t_{2,3})$, where $\t_{j,i}$ is the $j^{th}$ basis vector for the $i^{th}$ time period, with coefficient $c_{j,i}^t$, illustrated in Figure \ref{temporalbasis}.
\par
\begin{figure}[t]
\centering
\includegraphics[width=0.95\linewidth]{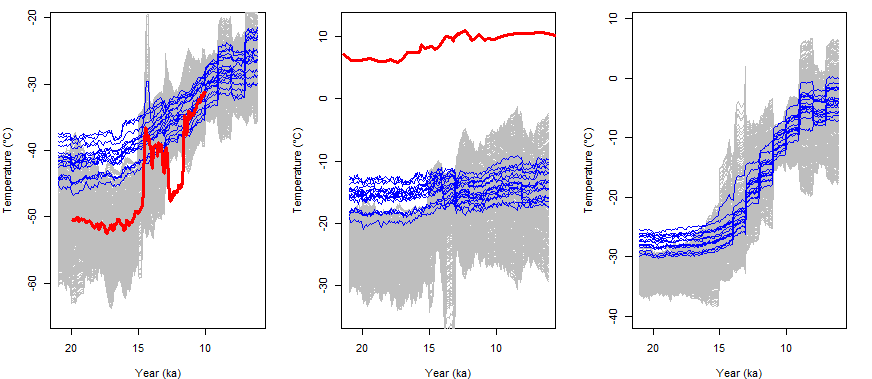}
\caption{The observed temperatures (red), climate ensemble (blue), and boundary conditions (grey) given by sampling 500 sets of coefficients for $\t$, in Greenland (left), Alaska (middle), and North America (right).}
\label{coeffonly}
\end{figure}

Figure \ref{coeffonly} shows the boundary condition for 500 settings of the coefficients if only the temporal basis vectors (and the mean $\bmu$) were used. In Greenland, the observations lie within the range of temperatures given by the 500 samples, as Greenland was used to select the basis vectors. In Alaska, however, a large bias between the observations and the modelled boundary condition remains, demonstrating the requirement of the second basis selection step. Alternatively, we could use Alaska together with Greenland when selecting $\t$, but this would result in a trade-off in the quality of fit for Greenland.
\par

\subsubsection{Fitting the spatial basis}

\begin{figure}[t]
\centering
\includegraphics[width=1\linewidth]{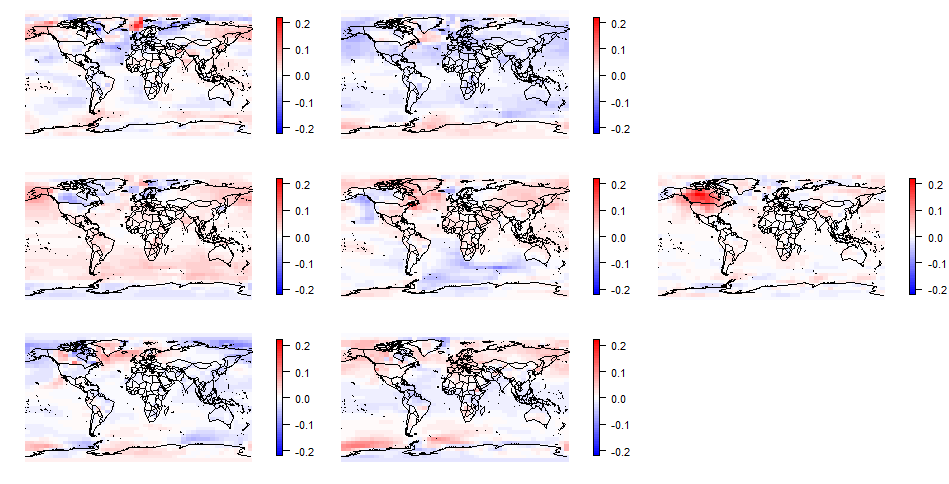}
\caption{Spatial basis vectors for the 3 time periods (top: 21 ka - 15 ka, middle: 15 ka - 13 ka, bottom: 13 ka - 6 ka), given using the method in Section \ref{spatialvectors}. Each vector is reasonably smooth spatially, and has some effect over North America.}
\label{spatialbasis}
\end{figure}

Given $\t$, we applied the method from Section \ref{spatialvectors} to select the spatial vectors, $\s$. We performed this separately for each time period, $i$, calculating $\epsilon_{\T}$ and $\epsilon_{\obs}$ as in \eqref{spatialresidual}, given the two temporal basis vectors already chosen, $\t_{1,i}, \t_{2,i}$, before finding a rotation, with $n_s = 2$ (chosen to trade-off minimising the number of coefficients in the model, and the error). The resulting spatial fields are shown in Figure \ref{spatialbasis}. Each of these patterns are reasonably smooth, and reduce the biases between the observations and the previously fitted components of the boundary condition model.
\par
When generating boundary conditions with this set of basis vectors, we found that we generally had extremely cold temperatures in the centre of North America in the second time period. Although there are no observations here, the ice sheet modellers in our team believed that the generated temperatures were too cold, and we found through test simulations on Glimmer that this led to much slower retreat of the ice sheet, compared to observations. We therefore added a third basis vector for the second time period, $\s_{3,2}$, allowing the temperature over North America to be increased, whilst still allowing the too cold temperatures if this coefficient is set close to 0 (with the goal that these temperatures would be ruled out when we history match due to leading to implausible ice sheets). This final basis vector, $\s_{3,2}$, is the far right plot in the second row of Figure \ref{spatialbasis}.
\par
In total, we have 13 coefficients that generate a full spatio-temporal boundary condition. 

\section{Prior boundary condition space} \label{definec}

For calibration, a bounded space of inputs is needed. We have ranges of possible settings for the ice sheet parameters, giving parameter space $\tspace$. We require a similar closed subset of $\mathbb{R}^{n_t + n_s}$, $\C$, so that $\c \in \C$. By varying $\c$ and evaluating $h(\c)$ in \eqref{boundaryModel}, we can compare a generated boundary condition to $\obs_{T, obs}$ via history matching.
\par
The `$j^{th}$ maximum implausibility measure' is \citep{craig1997pressure}:
\begin{displaymath}
\mathcal{I}_{jM}(\cdot) = \max_{i} (\{\mathcal{I}_{i}(\cdot)\} \backslash \{ \mathcal{I}_{M}(\cdot), \mathcal{I}_{2M}(\cdot), \ldots, \mathcal{I}_{(j-1)M}(\cdot) \}), \, \mathcal{I}_{M}(\cdot) = \max_{i} \mathcal{I}_i(\cdot).
\end{displaymath}
We calculate the implausibility (as in \eqref{impl}) of the prior boundary condition, $h(\cc)$, at each observed spatial location, $i$, with zero discrepancy and emulator variance, and observation error set according to the type of reconstruction (discussed in Section \ref{tempobs}), giving a set of implausibilities, $\{ \mathcal{I}_{i}(\cdot)\}$. Location $i$ has $\ell_i$ temporal observations, hence we standardise the implausibilities as:
\begin{equation} \label{scaledimpl}
\hat{\mathcal{I}}_{i}(\cdot) = 3\frac{\mathcal{I}_{i}(\cdot)}{b_i}, \quad b_i = \chi^2_{\ell_i, 0.995},
\end{equation}
and define coefficient space as:
\begin{displaymath}
\C = \{ \c \sim \pi(\cc) \, | \, \hat{\mathcal{I}}_{jM}(\c) < 3 \},
\end{displaymath}
ruling out coefficients, $\c$, sampled from prior $\pi(\cc)$ (bounds on $\cc$ set to, for example, avoid unnaturally extreme temperatures), if the boundary condition temperature for at least $j$ of the spatial locations is deemed to be implausible. This calculation is fast, as unlike standard applications of history matching (e.g. \citet{vernon2010galaxy,williamson2015bias}), an emulator is not required. 
\par
The resulting space, $\C$, contains values of $\cc$ that leads $h(\cc)$, the prior mean of the boundary condition $\T$, to be sufficiently similar to the temperature observations, so that we would wish to use $\T$ calculated from samples from $\C$ to force a deglaciation simulation with Glimmer. 
\par
Using the basis vectors selected in Section \ref{applicationsection}, we defined the coefficient space, $\C$, as above (shown in Figure \ref{prenroy}), sampling 500 values of $(\x, \c) \in \tspace\times\C$ using a Latin hypercube to give wave 1 design, $(\textbf{X} \times \textbf{C})^{(1)}$. We generated boundary condition $\T$ via \eqref{BCupdate}, and ran Glimmer with input $(\x, \T)$ to give wave 1 ensemble, $\ens^{(1)}$.
\par
\begin{figure}[t]
\centering
\includegraphics[width=1\linewidth]{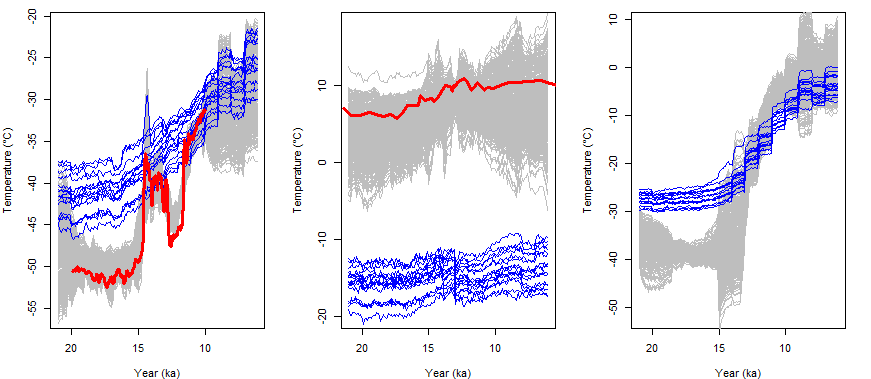}
\caption{The observed temperatures (red), climate ensemble (blue), and wave 1 (grey) boundary conditions, in Greenland (left), Alaska (middle), and North America (right).}
\label{BCplotw1}
\end{figure}

Figure \ref{BCplotw1} shows the boundary conditions given by $(\textbf{X} \times \textbf{C})^{(1)}$ (grey lines), for Greenland, Alaska, and the centre of North America. For Greenland and Alaska, our model produced temperatures more consistent with the observations (red) than the climate ensemble (blue) does, with the rapid warming and cooling periods captured. Initially, our model has colder temperatures than the climate ensemble in North America, but offers a wider range throughout, with the GCM runs contained within this spread from 14 ka onwards. This colder initial temperature may be accurate, given the climate ensemble is known to be too warm in Greenland. Overall, our model allows the boundary condition to be varied more than in any previous study.

\subsection{Calibrating $(\x, \cc)$}

Our aim is to calibrate $(\x, \cc)$ using ice sheet observations. Having seen $\obs_{T, obs}$, and used this to derive $\C$, we view $\pi(\c \mid \obs_{T, obs}) \sim Uniform(\c \in \C)$. Hence, our samples $\T \mid \obs_{T, obs}, \c$ (equation \eqref{BCupdate}) are samples from the joint distribution $\pi(\T, \c \mid \obs_{T, obs})$:
\begin{displaymath}
\pi(\T, \c \mid \obs_{T, obs}) = \pi(\T \mid \obs_{T, obs}, \c) \pi(\c \mid \obs_{T, obs}).
\end{displaymath}
Using the ice sheet observations, we history match $\c$ in order to further constrain our distribution for the full boundary condition $\T$, as the small number of parameters in $\c$, compared to the dimension of $\T$, makes this a manageable calibration problem when combining these parameters with $\x$.

\section{History matching to binary data} \label{binarysection}

We have spatial binary observations for the ice extent, $\obs^{ice}$, at certain time points, whereas Glimmer's output is ice thickness. \citet{chang2016calibrating} provide an extension to Bayesian calibration for binary data, although in their application the model output itself is binary. We choose to not treat the Glimmer output as binary, allowing the thickness to be incorporated into the emulation and calibration process rather than ignoring this information.
\par
\citet{gregoire2016abrupt} rule out runs by counting the number of boxes that are misclassified, based on setting a threshold for model thickness that can be treated as `no ice'. The number of misclassified boxes was calculated for only the observed runs of Glimmer. Here, we define a similar measure using the emulated thickness, allowing the entire input space to be searched, as is standard in history matching.

\subsection{General formulation}

We assume that binary observations, $\obs^b$, are the binary representation of the modelled latent process, $f(\x)$ (in our case, ice thickness), at its best input setting $\best$ (given observation error $\err$ and discrepancy $\disc$, as in Section \ref{sectionhm}):
\begin{displaymath}
\obs^b = \mathds{1}^b (f(\best) + \err + \disc),
\end{displaymath}
where a general $\ell$-dimensional spatial field $f(\x)$ is converted to binary via:
\begin{equation} \label{binaryeqn}
f^b(\x) = \mathds{1}^b (f(\x)), \quad [f^b(\x)]_i = \begin{cases}
0 \quad \text{if} \quad [f(\x)]_i \leq T^b \\
1 \quad \text{otherwise} \\
\end{cases} i = 1, \ldots, \ell,
\end{equation}
for a threshold of $T^b$. Given emulators for the spatial field, we account for the uncertainty in the binary representation at $\x$ by drawing $m$ samples from the emulator posterior, and converting each to binary via \eqref{binaryeqn}. We assess the distance between $\obs^b$ and sample $j$, $f^b_j(\x)$, via the number of misclassified grid boxes, as in the usual ice sheet modelling approach \citep{gregoire2016abrupt}:
\begin{displaymath}
\mathcal{I}_j^b(\x) = (\obs^b - f^b_j(\x))^T (\obs^b - f^b_j(\x)), \quad \mathcal{I}^b(\x) = (\mathcal{I}_1^b(\x), \ldots, \mathcal{I}_m^b(\x)),
\end{displaymath}
leading us to define NROY space as:
\begin{displaymath}
\mathcal{X}_{NROY} = \{ \x \in \mathcal{X} | P(\mathcal{I}^b(\x) \leq N_T) \geq 0.05 \},
\end{displaymath}
i.e. $\x$ is ruled out if it is unlikely that the number of misclassified grid boxes falls below a set threshold, $N_T$ (based on expert judgement as to what an acceptable misclassification rate is for a given field).
\par

\subsection{Application to Glimmer}

We spatially emulate the thickness by calculating the SVD basis of the centred ensemble (as in \eqref{svd}), finding an optimal rotation (equation \eqref{rotationminimise}, with $\weight$ the identity matrix), and then emulating the coefficients on this basis (\eqref{projecteqn} and \eqref{projectemulate}), as described in Section \ref{emulationhm}. The emulated coefficients are used to reconstruct fields of ice thickness, and in order to history match, we use the above method, with $\x$ replaced by $(\x, \c)$ as the set of calibration parameters.
\par
For setting the threshold, $N_T$, in this study we use a similar rule-of-thumb as in \citet{gregoire2016abrupt}, where simulations with an extent error of $>23\%$ were discarded, with this tolerance chosen based on comparing Glimmer output to observations, and judging which are acceptable. We use 25\% as a baseline for the acceptable error, although for the regions considered in Section \ref{hmresults}, this is not always suitable, as it can lead to ruling out either all, or none, of $\mathcal{X} \times \C$, hence we adjust $N_T$ in these cases. Additionally, instead of ruling out runs with a low probability of having an error below $N_T$, we summarise the posterior samples from the emulator by taking the minimum or mean error across $\mathcal{I}^b(\x)$, and comparing this value to the threshold (as in this application, this gave greater accuracy, in terms of ruling out ensemble members judged to be inconsistent with observations).

\section{History matching Glimmer} \label{hmresults}

We performed 3 waves of history matching, iteratively ruling out space that was inconsistent with ice sheet observations, using both the volume and extent of the ice sheet as a proxy for the thickness output of Glimmer. Instead of emulating the full spatio-temporal output, at each wave we selected different aspects to be emulated, and removed clearly unphysical behaviours (a benefit of history matching is that it does not require an accurate emulator for every output at once). Table \ref{hmtable} summarises the outputs emulated, at which waves, and the implausibility measures and bounds that were used. For the ice sheet volumes, the error variances were estimated from \citet{tarasov2012data}.

\begin{figure}[t]
\centering
\includegraphics[width=1\linewidth]{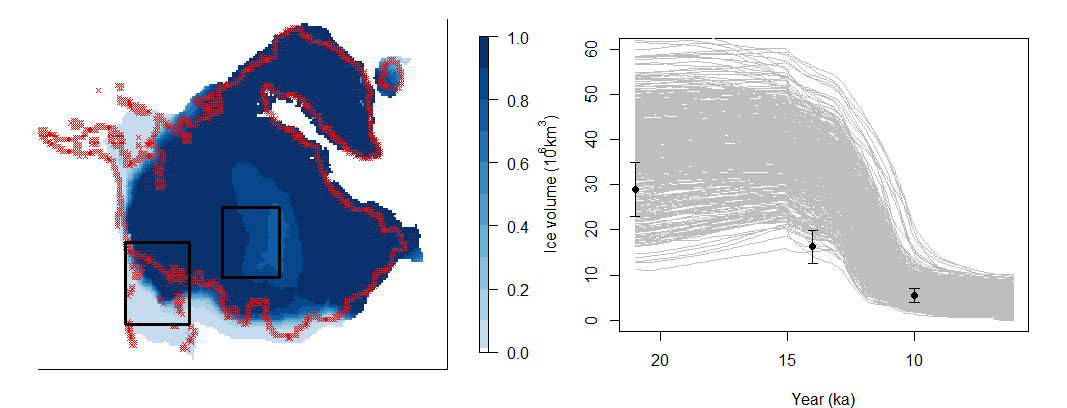}
\caption{Left: the proportion of wave 1 ensemble members containing ice in a grid box at 21 ka, with red crosses indicating the ice extent, and the wave 1 emulated regions shown by the two boxes. Right: the volume of wave 1 ensemble members in time, with observations (with error) shown in black.}
\label{wave1results}
\end{figure}

\subsection{Wave 1} \label{sectionw1}

At wave 1, only 21 ka was considered, as the output for later years is dependent on the initial ice sheet, and due to the wide range of behaviours possible by varying the boundary condition coefficients, the initial ice sheet is often clearly implausible. Figure \ref{wave1results} shows the average ice extent at 21 ka across the wave 1 ensemble, and the volume through time, with a large spread of potential volumes at 21 ka. We built emulators for the volume at 21 ka, and two separate regions of spatial output, indicated in Figure \ref{wave1results}, covering regions where there was a difference between the ensemble and the extent reconstruction at 21 ka: the centre of the ice sheet, where there should be ice, and the south-west, where Glimmer's ice coverage generally extends too far.
\par
When emulating spatial regions, we followed the rotation and emulation methodology outlined in Section \ref{emulationhm}. For the volume at a single time point, the emulator was a stationary Gaussian process. For the implausibility in a spatial region, we used the method from Section \ref{binarysection}, with $T^b = 10$ throughout. For the central region, the mean of $\fimpl^b(\x, \c)$ ruled out space accurately when considering the known ensemble runs (using the $5^{th}$ percentile failed to rule out several runs that poorly matched $\obs^{ice}$, as the distribution of $\fimpl^b$ was often bimodal for this region when the latent process was close to $T^b$), whereas $min(\fimpl^b(\x,\c))$ was more accurate for the south-west region.
\par
Wave 1 NROY space is defined as:
\begin{displaymath}
(\tspace \times \C)_{NROY}^{(1)} = \{ (\x, \c) \in \mathcal{X}\times\C \, | \, \mathcal{I}_i(\x, \c) < b_i, i = (vol21, sw21, ce21) \},
\end{displaymath}
with the bounds $b_i$ given in Table \ref{hmtable}. $(\tspace \times \C)_{NROY}^{(1)}$ consists of 5.4\% of the original space, and is illustrated in Figure \ref{w1nroy}. We sampled 500 points from $(\tspace \times \C)_{NROY}^{(1)}$, chosen via a combination of space-filling and including parameter settings leading to the lowest implausibilities for each emulated output, and ran Glimmer to obtain the wave 2 ensemble. Figure \ref{wave2results} compares waves 1 and 2.

\subsection{Waves 2 and 3} \label{sectionw23}

At wave 2, we used metrics throughout the deglaciation (21 ka, 14 ka, 10 ka, 6 ka), allowing coefficients controlling the boundary condition in each of the three time periods to be constrained. These time points were chosen based on discussions with experts, and to give a coverage of the entire deglaciation. The spatial regions were chosen by selecting regions towards the edge of the ice sheet, where the ensemble generally does not match observations, but where there is a range of behaviours in the ensemble (left halves of Figures \ref{w23pro14ka}, \ref{w23pro10ka}, \ref{w23pro6ka}).
\par

\begin{table}[t]
\centering
\begin{tabular}{|c|c|c|c|c|c|c|}
\hline
\textbf{Output} & $id$ & \textbf{Waves} & \textbf{Impl} & $\ell$ & $\sigma^2_{\err}$ & $b_{id}$ \\ \hline
21 ka volume & $vol21$ & 1,2 & $\fimpl$ & 1 & 4 & $3^2$ \\  \hline
21 ka south-west region & $sw21$ & 1 & $\fimpl^b$ & 1116 & & 0.25$\ell$  \\ \hline
21 ka central region & $ce21$ & 1 & $\fimpl^b$ & 868 & & 0.025$\ell$  \\ \hline
14 ka volume & $vol14$ & 2,3 & $\fimpl$ & 1 & 1.416 & $3^2$ \\ \hline
14 ka region & $reg14$ & 2 & $\fimpl^b$ & 1176 & & 0.33$\ell$  \\  \hline
10 ka volume & $vol10$ & 2 & $\fimpl$ & 1 & 0.279 & $3^2$ \\ \hline 
10 ka region & $reg10$ & 2 & $\fimpl^b$ & 1066 & & 0.25$\ell$  \\ \hline
6 ka region & $reg6$ & 2 & $\fimpl^b$ & 1271 & & 0.25$\ell$ \\ \hline
\end{tabular}
\caption{Information for outputs used in history matching. `Waves' indicates at which waves the output was emulated, `Impl' gives which implausibility was used, $\ell$ is the dimension of the output, $\sigma^2_{\err}$ is the observation error variance for the volumes, and $b_{id}$ is the history matching bound for output `$id$'.}
\label{hmtable}
\end{table}

Based on the 6 emulated outputs, we defined NROY space as (with $\hat{\mathcal{I}}$ the scaled implausibility as in \eqref{scaledimpl}):
\begin{displaymath}
(\tspace \times \C)_{NROY}^{(2)} = \{ (\x, \c) \in (\tspace \times \C)_{NROY}^{(1)} \, | \, \hat{\mathcal{I}}_{4M}(\x, \c) < 3 \},
\end{displaymath}
i.e. not requiring every output to be consistent with the observations, to reduce the risk of incorrectly ruling out good values of $(\x, \c)$. After wave 2, NROY space is 1.1\% of the full space (Figure \ref{w2nroy}). We obtain a wave 3 ensemble by sampling from $(\tspace \times \C)_{NROY}^{(2)}$ using the same considerations as at the previous wave. Figures \ref{w23pro14ka}, \ref{w23pro10ka} and \ref{w23pro6ka} show how the ice extent has changed between waves 2 and 3 for the emulated regions.
\par
With no more resources for running Glimmer, we used the volume at 14 ka to give a final NROY space, as this was a difficult constraint to satisfy. Emulating using the wave 3 ensemble, the wave 3 NROY space is:
\begin{displaymath}
(\tspace \times \C)_{NROY}^{(3)} = \{ (\x, \c) \in (\tspace \times \C)_{NROY}^{(2)} \, | \, \mathcal{I}_{vol14}(\x, \c) < 3^2 \},
\end{displaymath}
and consists of 0.06\% of the original space (Figure \ref{w3nroy}).

\subsection{Boundary condition uncertainty}

Our initial goal was to calibrate the boundary condition of Glimmer, hence we now consider how the space of possible boundary conditions has evolved due to history matching.

\begin{figure}[t]
\centering
\includegraphics[width=1\linewidth]{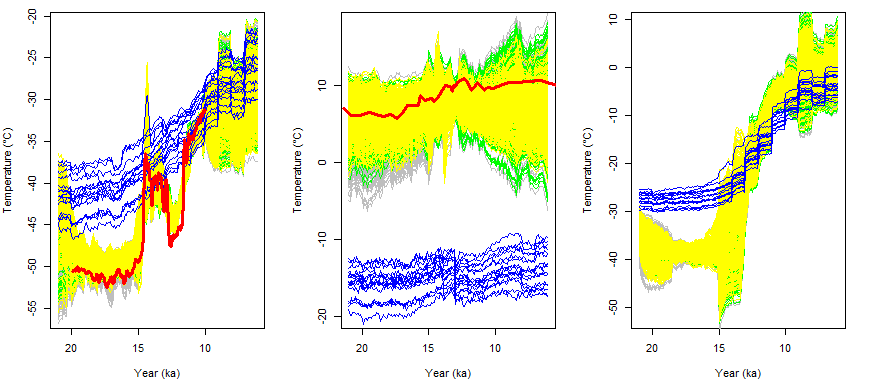}
\caption{The observed temperatures (red), climate ensemble (blue), wave 1 (grey), wave 2 (green) and wave 3 (yellow) boundary conditions, in Greenland (left), Alaska (middle), and North America (right).}
\label{BCplot}
\end{figure}

Figure \ref{BCplot} shows the simulated boundary conditions for each of the three ensembles, compared to the observations and climate ensemble. In each location, we have started to reduce the spread of plausible temperatures, e.g. for each, we have ruled out the coldest temperatures in the first time period. Even though the original dimension of the input was over 300 million, we can quantify and constrain uncertainty in the boundary condition using only 13 coefficients to represent it.
\par
\begin{figure}[t]
\centering
\includegraphics[width=1\linewidth]{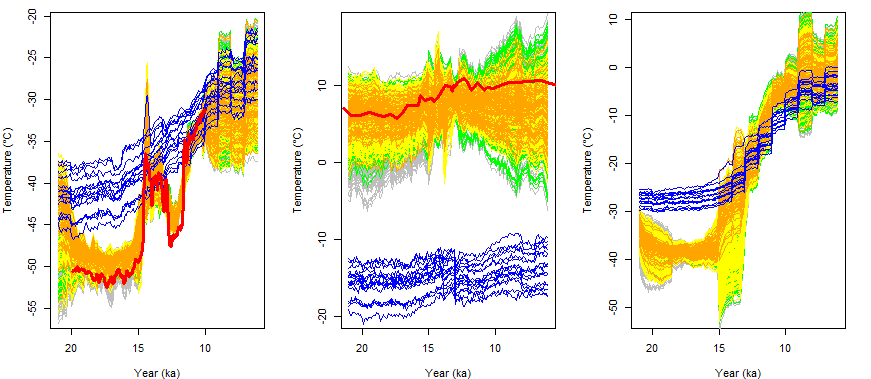}
\caption{The observed temperatures (red), climate ensemble (blue), wave 1 (grey), wave 2 (green), wave 3 (yellow) and the not ruled out wave 3 (orange) boundary conditions, in Greenland (left), Alaska (middle), and North America (right).}
\label{BCplot4}
\end{figure}

Figure \ref{BCplot4} shows the spread of boundary conditions for the wave 3 ensemble members that were not ruled out after wave 3 (orange lines), compared to the previous waves. When we only consider these runs, we see that the range of boundary condition temperatures that are possible in each location has been reduced more substantially than at previous waves. In Greenland, the warmest peaks around 13 ka have been ruled out, as well as the coldest initial temperatures. In both Alaska and North America, the spread of temperatures in the first time period has been reduced further. All of the boundary conditions featuring the large, unrealistic, downward temperature shifts in North America have been ruled out at wave 3, leaving temperature profiles that generally increase through time, as expected. History matching has enabled us to rule out clearly unphysical boundary conditions.
\par
The final range of temperatures in North America (over the ice sheet) has been reduced from the initial space of boundary conditions that we allowed. The range of not implausible boundary conditions is similar to the range given by the climate ensemble, but with a different warming trajectory. At the start of the deglaciation, our boundary condition allows colder temperatures. At the end, we initially had a spread of possible temperatures covering the climate ensemble, but ruled out most of the colder boundary conditions, suggesting that the climate ensemble may be too cold after 10 ka.

\subsection{Ice sheet volume}

\begin{figure}[t]
\centering
\includegraphics[width=1\linewidth]{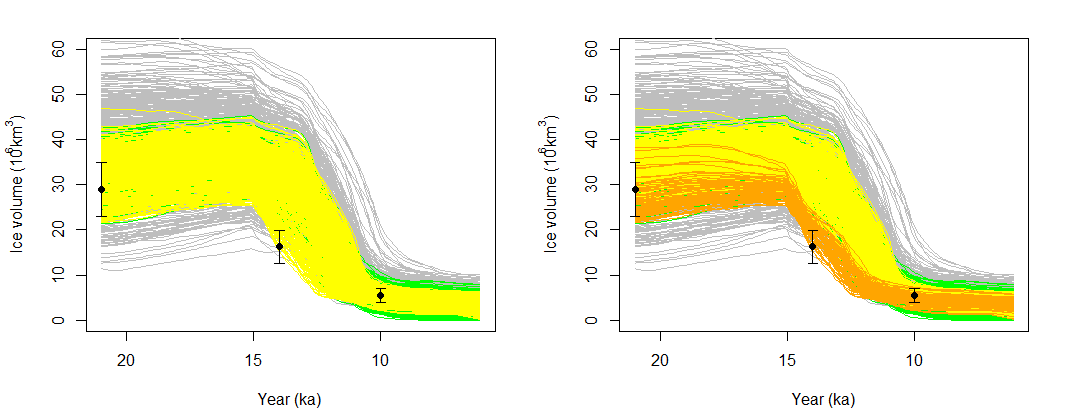}
\caption{Ice sheet volume for the wave 1 (grey), wave 2 (green), and wave 3 (yellow) ensembles, with the observations and observational error shown in black. The orange runs added in the right plot are those not ruled out after wave 3.}
\label{wave3results}
\end{figure}

Figure \ref{wave3results} shows how the ice volume evolves through time, for the wave 1 ensemble (grey lines), wave 2 (green), and wave 3 (yellow).
\par
At wave 1, there is a wide range of behaviours, especially at 21 ka, as we used a space-filling design of $\tspace\times\C$. The waves 2 and 3 ensembles generally look similar, with the most extreme runs from wave 1 ruled out, and with the observations at 21 ka and 10 ka often being matched. Few runs in the wave 2 ensemble matched the volume observation at 14 ka, with the majority of ice sheets retreating too slowly. This problem was sometimes fixed in the wave 3 ensemble, with the volumes from 10 ka onwards also in a narrower, more accurate, range.
\par
As the wave 3 NROY space was defined using only an emulator for the volume at 14 ka, the ensemble members that are not ruled out (coloured orange in the right of Figure \ref{wave3results}), generally match the observed time series of ice melt. This, combined with Figure \ref{BCplot4}, validates the method: we have found a subset of the initial boundary condition space, distinct from the original climate ensemble, that allows Glimmer to reproduce the ice sheet volume through the deglaciation.

\section{Discussion} \label{discussion}

We have developed a framework for calibrating high-dimensional boundary conditions that have sparse observational data. Our method allows a range of plausible boundary conditions to be efficiently generated, given a small ensemble of climate model runs and sparse observations. The parameterised low-dimensional form we used allows calibration of the boundary conditions and computer model parameters to be performed jointly, quantifying the uncertainty in the output due to each. Our method allows both a wider range of, and more historically accurate, boundary conditions to be used.
\par
Performing three waves of history matching on the Glimmer ice sheet model has demonstrated that it is possible to calibrate the high-dimensional boundary conditions of complex computer models, with the range of possible temperatures reduced, whilst also allowing the interaction of the boundary conditions with the ice sheet parameters to be explored. The reduction in the spread of ice sheet volumes, and the improvement in the ice extents, shows the success of this method. At wave 3, we have started to identify runs with a melting trajectory consistent with observations of the deglaciation. With further ensembles of Glimmer, we would explore this space to identify runs with ice extents that are more consistent with observations through the deglaciation.
\par
Our method for history matching binary observations successfully identified runs with ice extent more consistent with the observations than in the initial ensemble, although there was still a reasonably large discrepancy between the wave 3 ensemble and the truth. To identify whether this is due to not yet finding the region of the input space that leads to output most consistent with ice extent reconstructions, or that this is structural error, would require further waves of history matching. Modelling the precipitation boundary condition may give further improvements, although this problem suffers from an even greater lack of geological observations.
\par
A future extension of the binary history matching method presented here would be to consider the probability that the latent thickness process matches the observations in each grid box, rather than just using the fixed threshold to convert ice thickness to binary. This would likely improve matching in regions where the emulated thickness is relatively low, and hence close to the threshold for the presence of ice.
\par
Enabling model performance to be explored under a range of realistic boundary conditions may lead to improved future development of ice sheet models, with a better understanding of which processes cannot be represented currently, and where discrepancies with observations lie. In general, our modelling framework can be combined with expert elicitation, for example to explore the effect that certain patterns or changes to the boundary conditions have on the ice sheet.
\par
Ultimately, improved ice sheet models that more accurately simulate the past will help with understanding the deglaciation. Reducing the uncertainty in the past temperatures required to give a realistic deglaciation may eventually help to improve future projections, with these being used as out-of-sample constraints for the output of global climate models run into the future.

\bibliographystyle{apa}
\bibliography{fullbib}

\newpage

\setcounter{equation}{0}
\setcounter{figure}{0}
\setcounter{table}{0}
\setcounter{page}{1}
\setcounter{section}{0}
\makeatletter
\renewcommand{\thefigure}{S\arabic{figure}}
\renewcommand{\thesection}{S\arabic{section}}
\renewcommand{\thetable}{S\arabic{table}}
\renewcommand{\theequation}{S\arabic{equation}}

\section{Calculations} \label{smcalculation}

In Section \ref{completeprocess}, we use the expectation of $\pi(\T \mid \obs_{T, obs}, \c)$. From \eqref{boundaryModel} and \eqref{obsModel}, we have:
\begin{displaymath}
\var_e = \var_s\otimes\var_{t'}, \quad \var_{\eps} = \var_s\otimes\var_{t}.
\end{displaymath}
In \eqref{BCupdate}, we need $\pi(\obs | \cc)$, which can be written as:
\begin{displaymath}
\pi(\obs | \cc) \propto \int \pi(\obs | \T, \cc) \pi(\T | \cc) d\T,
\end{displaymath}
with the two components of this expression given by equations \eqref{obsModel} and \eqref{boundaryModel} respectively. Using that both have multivariate Normal distributions, we have:
\begin{displaymath}
\obs | \cc \sim \mathrm{MVN}(h(\cc), (\var_{\err}^{-1} - \var_{\err}^{-1} (\var_{\err}^{-1} + \var_{\eps}^{-1})^{-1} \var_{\err}^{-1})^{-1}).
\end{displaymath}
We sample a full observation vector, $\obs_T$, for the spatial locations where we have some observations, using the Normality of  $\obs | \cc$. Then, by conditioning on $\obs_T$, we have:
\begin{align*}
\begin{split}
\text{E}[\T \mid \obs_T, \cc] &= h(\cc) + (\var_{s} \otimes \var_t) [(\var_{s} \otimes \var_{t'}) + (\var_{s} \otimes \var_t)]^{-1} vec((\obs_T - h(\cc))^T) \\
&= h(\cc) + (\var_{s} \otimes \var_t) [(\var_{s} \otimes (\var_{t'} + \var_t)]^{-1} vec((\obs_T - h(\cc))^T) \\
&= h(\cc) + (\var_{s} \otimes \var_t) (\var_{s}^{-1} \otimes (\var_{t'} + \var_t)^{-1}) vec((\obs_T - h(\cc))^T) \\
&= h(\cc) + (\var_{s} \var_{s}^{-1} \otimes (\var_t (\var_{t'} + \var_t)^{-1}) vec((\obs_T - h(\cc))^T) \\
&= h(\cc) + vec(\var_t (\var_{t'} + \var_t)^{-1} (\obs_T - h(\cc))^T (\var_{s} \var_{s}^{-1})^T).
\end{split}
\end{align*}
If there was not either a common spatial or temporal variance matrix, this calculation would require expanding the Kronecker product, which is not tractable due to the high dimension of the problem.

\subsection{Temporal vectors} \label{smtemporal}

Rewriting equation \eqref{svdbasis}, we see that the SVD basis vectors, and the rotated basis, are linear combinations of the (centred) ensemble members:
\begin{displaymath}
\Tau_{\bmu, \mathcal{S}_t} (\textbf{U} \boldsymbol{\Sigma}^{-1}) = \bas, \quad \Tau_{\bmu, \mathcal{S}_t} (\textbf{U} \boldsymbol{\Sigma}^{-1} \rot) = \bas \rot.
\end{displaymath}
Therefore, a linear combination of the $n_t$ basis vectors can be rewritten as a linear combination of the (centred) ensemble:
\begin{align*}
\begin{split}
\sum_{j = 1}^{n_t} c_j^t [\bas \rot]_{\cdot j} &= \sum_{j = 1}^{n_t} c_j^t [\Tau_{\bmu, \mathcal{S}_t} (\textbf{U} \boldsymbol{\Sigma}^{-1} \rot)]_{\cdot j} \\
&= \sum_{j = 1}^{n_t} c_j^t \sum_{k = 1}^n \tau_{\bmu, \mathcal{S}_t}(\x_k) [(\textbf{U} \boldsymbol{\Sigma}^{-1} \rot)]_{kj} \\
&= \sum_{k = 1}^n ( \sum_{j = 1}^{n_t} c_j^t [(\textbf{U} \boldsymbol{\Sigma}^{-1} \rot)]_{kj}) \tau_{\bmu, \mathcal{S}_t}(\x_k).
\end{split} 
\end{align*}
By removing the index $\mathcal{S}_t$, we have coefficients defining a full spatio-temporal field, a linear combination of $\Tau_{\bmu}$, and hence the temporal basis vectors are given by
\begin{displaymath}
\t_j = [\Tau_{\mu} (\textbf{U} \boldsymbol{\Sigma}^{-1} \rot)]_{\cdot j}, \quad j = 1, \ldots, n_t.
\end{displaymath}

\subsection{Smoothing} \label{smoothing}

The initial constraints on $h(\c)$ remove very unphysical behaviours. To also reduce jumps in the temperature between time periods, we applied a smoothing between the second and third time periods (ignoring the previous transition as we did not want to remove the observed jump in Greenland, close to the boundary between the first two periods). When history matching with Glimmer, we find whether we can remove boundary conditions with such jumps between the first and second time periods from the plausible set. To smooth, we set:
\begin{displaymath}
h(\cc)_t = \frac{1}{7} \sum_{i = t - 3}^{t + 3} h(\cc)_{i}, \quad t = 78, \ldots, 83,
\end{displaymath}
where $h(\cc)_i$ gives the entries of $h(\cc)$ at time $i$, and where we have used 100 year averages as our time points, so that $t = 78, \ldots, 83$ relates to 13.3 ka to 12.8 ka.

\subsection{Seasonality} \label{seasonality}

To run Glimmer, we require monthly fields for each 100 years, rather than the mean for the whole 100 years. However, the terms that control the temporal variability in $h(\c)$ (mean $\bmu$, basis vectors $\textbf{t}_j$) can be converted from 100 year annual averages to 100 year averages for individual months, without affecting the match that was achieved for the observed temperatures. This is done similarly as when converting the temporal basis vectors from a subset of locations to all locations in Section \ref{temporalvectors}, as we have monthly climate model output. The spatially-derived basis vectors remain unchanged, as they do not vary in time.
\par
Let a subscript of $t$ denote the time point, so that
\begin{displaymath}
h(\c)_t = \bmu_t + \sum_{j=1}^{n_t}c^t_j\t_{jt} + \sum_{j=1}^{n_s}c^s_j\s_{jt},
\end{displaymath}
gives the spatial field for time $t$. In the application, an individual value of $t$ represents a 100 year period. Then we have:
\begin{displaymath}
\bmu_{t} = \frac{1}{12} \sum_{m=1}^{12} \bmu_{tm},
\end{displaymath}
where a subscript of $m$ denotes the month, i.e. the ensemble mean for 100 year period, $t$, is the mean of the monthly ensemble means over the same 100 year period. Similarly, as $\t_{jt}$ is a linear combination of the (centred) ensemble at time $t$, we also have:
\begin{displaymath}
\t_{jt} = \frac{1}{12} \sum_{m=1}^{12} \t_{jtm}.
\end{displaymath}
This is given by decomposing the centred ensemble at time $t$, $\Tau_{\mu t}$, into its months, $\Tau_{\mu tm}, m = 1, \ldots, 12$, so that $\t_{jtm} = [\Tau_{\mu tm} (\textbf{U} \boldsymbol{\Sigma}^{-1} \rot)]_{\cdot j}$. Therefore, the input for time $t$ and month $m$ is given by:
\begin{displaymath}
h(\c)_{tm} = \bmu_{tm} + \sum_{j=1}^{n_t}c^t_j\t_{jtm} + \sum_{j=1}^{n_s}c^s_j\s_{jt},
\end{displaymath}
and because:
\begin{displaymath}
h(\c)_t = \frac{1}{12} \sum_{m=1}^{12} h(\c)_{tm},
\end{displaymath}
we still have the same mean temperature at time $t$, with monthly fields that can be used as inputs for Glimmer.

\begin{figure}[t!]
\centering
\includegraphics[width = 0.7\linewidth]{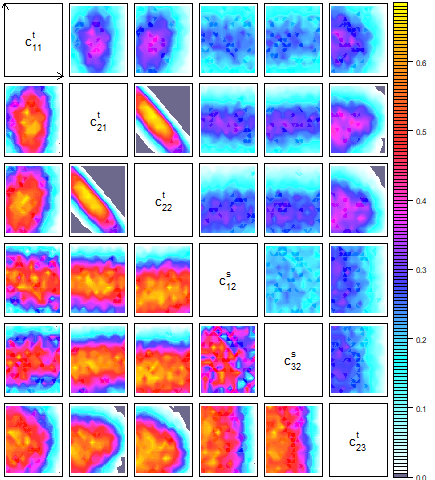}
\caption{The coefficient space, $\C$, for 6 of the coefficients. In each panel, we sample from the remaining dimensions and calculate the proportion of space that is not ruled out. The lower half plots have colour scales set individually to accentuate the relationships in each panel, with the axes aligned as in the top right plots.}
\label{prenroy}
\end{figure}

\begin{figure}[t!]
\centering
\includegraphics[width = 0.7\linewidth]{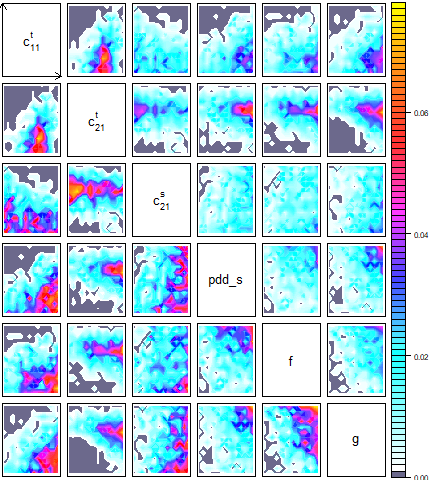}
\caption{The wave 1 NROY space for 3 of the coefficients and 3 ice sheet parameters. The lower half plots have colour scales set individually as before.}
\label{w1nroy}
\end{figure}

\begin{figure}[t!]
\centering
\includegraphics[width=1\linewidth]{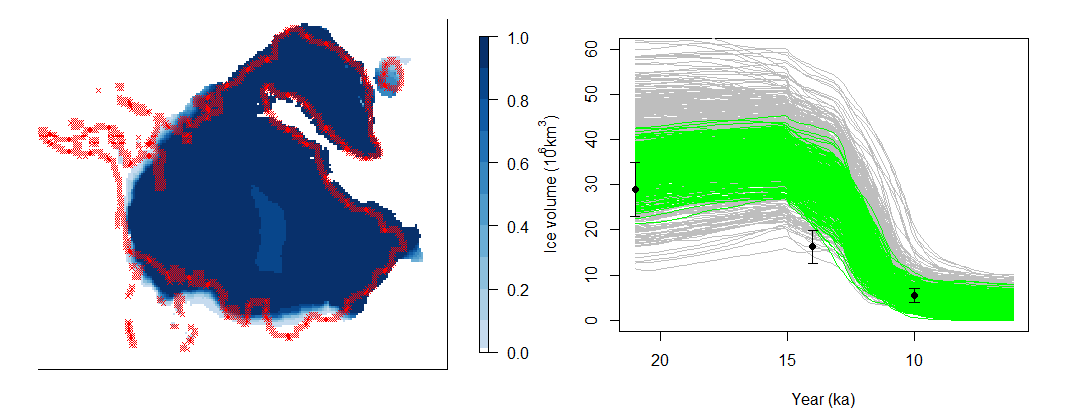}
\caption{The proportion of grid boxes in the wave 2 ensemble containing ice at 21 ka (left), and the volume of the wave 2 ensemble members (green) compared to wave 1 (grey).}
\label{wave2results}
\end{figure}

\subsection{Wave 1}

At wave 1, we only history matched to observations at 21 ka. Sampling from the full parameter-coefficient space leads to a wide range of ice sheet behaviours, the majority of which will not be realistic. By initially only reducing space based on implausible output at 21 ka, the wave 2 output should be more consistent with the observations at the start of the deglaciation, leading to a wider range of plausible deglaciations than at wave 1: if the ice sheet is not correct at 21 ka, we are unlikely to learn anything at subsequent times.
\par
It would have been possible to emulate the entire ($194\times150$) spatial output at 21 ka using the low-dimensional basis emulation approach described in Section \ref{emulationhm}, but we instead selected spatial regions where there were differences between the model output and observations, so that emulation is more efficient. If we never have ice in a certain grid box in the ensemble, then the basis we construct will also never have ice there. Therefore, we restricted emulation to spatial regions where there was a range of behaviours across the ensemble, with the dimension of the output, $\ell$, around 1000.
\par
The left half of Figure \ref{wave1results} shows the proportion of wave 1 ensemble members that contain ice in a given grid box at 21 ka, compared to the observed extent of the ice sheet (red crosses). The main differences are along the southern edge, where the ensemble has often extended too far; in the north-west, where the ensemble never has ice in Alaska; and in the centre, where some ensemble members do not have ice. The right panel of Figure \ref{wave1results} shows the volume for each ensemble member, with a wide range of starting volumes observed. 
\par
See Section \ref{sectionw1} for emulation and history matching details.

\begin{table}[t]
\centering
\begin{tabular}{|c|c|c|c|}
\hline
\textbf{Region} & \textbf{Wave 1} & \textbf{Wave 2} & \textbf{Wave 3} \\ \hline
21 ka (SW) & 214 (316.4) & 207 (261.5) &   \\  \hline
21 ka (central) & 0 (108.5) & 0 (53.5) &  \\ \hline
14 ka &  & 326 (483.3) & 254 (476.4) \\  \hline
10 ka &  & 163 (582.5) & 149 (451.1)  \\ \hline
6 ka &  & 108 (339.7) & 106 (311.9) \\ \hline
\end{tabular}
\caption{The minimum number of incorrect boxes for each region for each ensemble, with the average across the ensemble in brackets.}
\label{wrongtable}
\end{table}

\subsection{Wave 2}

After running the wave 2 design, $(\textbf{X} \times \textbf{C})^{(2)}$, on Glimmer, we checked whether we had improved the model output compared to wave 1. Figure \ref{wave2results} shows the ice extent in the wave 2 ensemble, showing that runs where there was significantly too much ice at the south-west edge of the ice sheet in wave 1 (Figure \ref{wave1results}) have now been ruled out, so that the output is more consistent with the observed extent at 21 ka. There are also fewer runs in the wave 2 ensemble with an opening in the centre of the ice sheet, although all parameter settings that lead to this have not yet been ruled out. The right half of Figure \ref{wave2results} shows that we have ruled out runs with a significantly too high or low volume at 21 ka, with the spread of possible volumes at wave 2 (green) much smaller than at wave 1.
\par
Table \ref{wrongtable} gives the minimum and average number of incorrect grid boxes for each region at waves 1 and 2, with the wave 2 ensemble generally closer to the observations, and with a better run having been found.
\par
\begin{figure}[t]
\centering
\includegraphics[width = 0.9\linewidth]{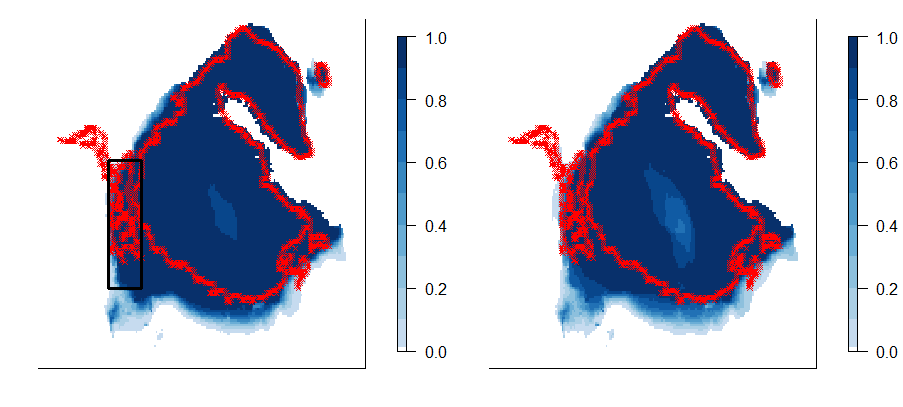}
\caption{The proportion of runs in the wave 2 (left) and wave 3 (right) ensembles with ice in each grid box at 14 ka, with the extent given by the crosses, and the emulated region by the box. There is not much of a difference between the ensembles in terms of the presence of ice in the chosen region, though elsewhere there are now ensemble members that are closer to the correct extent in the south (although these have still not retreated enough).}
\label{w23pro14ka}
\end{figure}
From the right panel of Figure \ref{wave2results}, we see that we don't currently match the observed volume at 14 ka: every run in the wave 2 ensemble has an ice sheet that melts too slowly. The spread of volumes at 10 ka is also too wide, although a number of runs do match the observations here. This suggests that the deglaciation generally starts too late in the wave 2 ensemble, but once the ice starts to melt, it retreats rapidly enough.
\par
\begin{figure}[t]
\centering
\includegraphics[width = 0.9\linewidth]{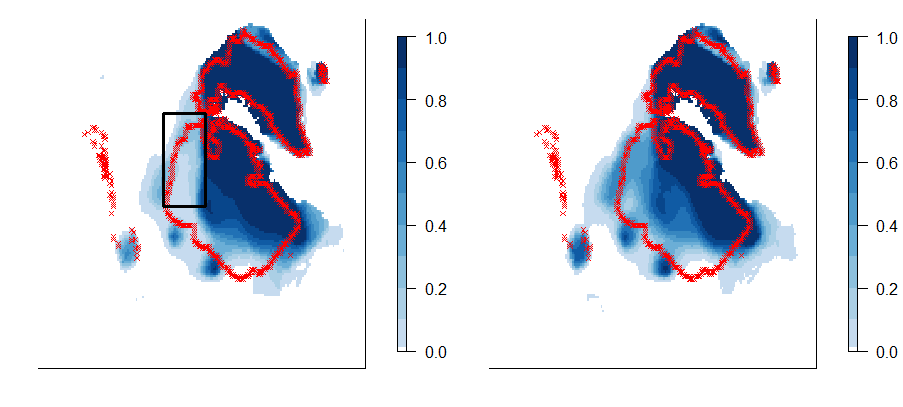}
\caption{The proportion of runs in the wave 2 (left) and wave 3 (right) ensembles with ice in each grid box at 10 ka, with the extent given by the crosses, and the emulated region by the box. At wave 3, there are more runs that have ice in the chosen region, as required, but there are still ensemble members that extend too far to the west.}
\label{w23pro10ka}
\end{figure}
\begin{figure}[t]
\centering
\includegraphics[width = 0.9\linewidth]{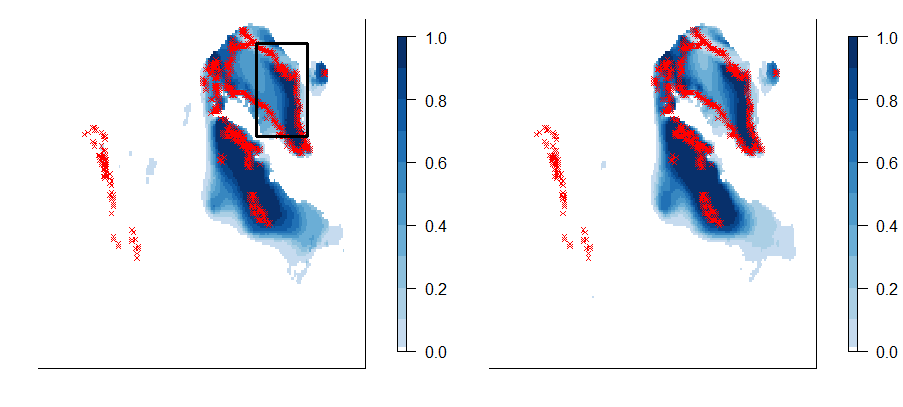}
\caption{The proportion of runs in the wave 2 (left) and wave 3 (right) ensembles with ice in each grid box at 6 ka, with the extent given by the crosses, and the emulated region by the box. Between waves 2 and 3, we have slightly improved the match in the region, although generally there is still too little ice in Greenland, and too much ice over the North American continent. There is now less ice in the west.}
\label{w23pro6ka}
\end{figure}

To search for runs that correct this, at wave 2 we built univariate emulators for the volume at each of 21 ka, 14 ka and 10 ka. We also selected spatial regions at each of 14 ka, 10 ka and 6 ka (left halves of Figures \ref{w23pro14ka}, \ref{w23pro10ka} and \ref{w23pro6ka}), with similar reasoning to the choice of regions at wave 1: we selected a subset of the output towards the current edge of the ice sheet, where we observed a variety of behaviours in the ensemble, allowing extremely bad runs to be ruled out, and so that we should find runs better than those we currently have. Emulating output from 14 ka onwards at this wave allows settings of the coefficients controlling the second and third time periods to be constrained, instead of only the first (as at wave 1).
\par

\begin{figure}[t]
\centering
\includegraphics[width = 0.7\linewidth]{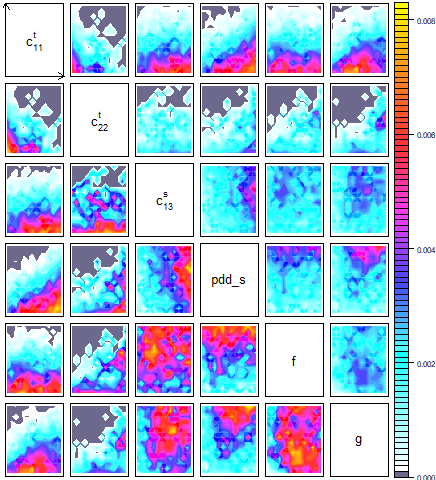}
\caption{The wave 2 NROY space for 3 of the coefficients and 3 ice sheet parameters. The lower half plots have colour scales set individually as before.}
\label{w2nroy}
\end{figure}

Figure \ref{w2nroy} illustrates the wave 2 NROY space, for one coefficient from each time period, and three of the ice sheet parameters.

\subsection{Wave 3}

We designed a final 500 member ensemble by sampling from within the wave 2 NROY space, $(\tspace \times \C)_{NROY}^{(2)}$. As before, we included runs that minimised the implausibility for each of the emulated outputs, and also selected 250 under a stricter definition of NROY space (settings for which at least 4, rather than 3, of the 6 outputs were not implausible): this was the final chance to run Glimmer, so we ran the model at parameter settings more likely to lead to realistic deglaciations. The remainder of the design was space-filling in $(\tspace \times \C)_{NROY}^{(2)}$, to give design $(\textbf{X} \times \textbf{C})^{(3)}$, and we evaluated Glimmer at these 500 settings.
\par
Figures \ref{w23pro14ka}, \ref{w23pro10ka} and \ref{w23pro6ka}, and Table \ref{wrongtable}, compare the output at waves 2 and 3, showing that better runs were found for each of the regions, in terms of reducing the number of grid boxes that are incorrect, whilst the whole ensemble is generally closer to the observations.
\par
\begin{figure}[t]
\centering
\includegraphics[width = 0.7\linewidth]{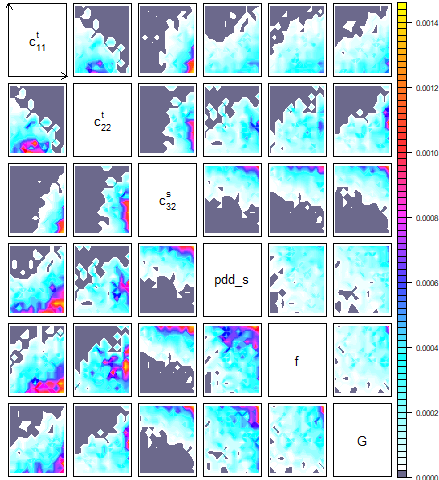}
\caption{The wave 3 NROY space for three coefficients and three ice sheet parameters. In each panel, we sample across the remaining inputs to calculate the proportion of space that is not ruled out. The lower left panels have colour scales set for each pairwise plot to show relationships within each panel, with orientation consistent with the upper right plots.}
\label{w3nroy}
\end{figure}

Matching to only the volume at 14 ka (Section \ref{sectionw23}), the wave 3 NROY space consists of 0.06\% of the original space. Figure \ref{w3nroy} highlights relationships between three ice sheet parameters and three boundary condition parameters, showing that we have ruled out space based on both types of inputs. The 3rd spatial coefficient for the 2nd time period, $c^s_{3,2}$, generally needs to have a high value, to achieve more accurate deglaciations, whilst the two temporal coefficients shown here can take on a wider range of values. The regions of the input space with the highest density of points in the current NROY space tend to be towards the edges of the allowed ranges, with ice sheet parameters $f$ (flow factor) and $g$ (geothermal heat flux) both needing high values.

\end{document}